\documentclass[anonymous=false, %
               format=acmsmall, %
               review=false, %
               screen=true, %
               nonacm=true]{acmart}

\setcopyright{none}
\copyrightyear{2021}

\usepackage{booktabs}       % professional-quality tables
\usepackage{amsmath}        % math/equations
\usepackage{mathtools}      % more math symbols like \coloneqq
\usepackage{amsfonts}       % blackboard math symbols
\usepackage{microtype}      % microtypography
\usepackage{diagbox}        % diagonal boxes at top of tabular

\usepackage{natbib}         % citations inline
\setcitestyle{round,authoryear,semicolon}

\usepackage[frozencache,cachedir=.]{minted}    % frozen styles for arXiv

\begin{document}

\title[Prioritizing municipal lead mitigation projects]{Prioritizing municipal lead mitigation projects as a relaxed knapsack optimization: a method and case study}

\author{Isaac Slavitt}
%\authornote{http://www.isaacslavitt.com}
\email{slavitt@seas.harvard.edu}
\orcid{0000-0002-7418-9876}
\affiliation{%
  \institution{Institute for Applied Computational Science, Harvard University}
  \streetaddress{150 Western Ave, Suite 1-312}
  \city{Boston}
  \state{Massachusetts}
  \country{USA}
  \postcode{02134}
}

%%%%%%%%%%%%%%%%%%%%%%%%%%%%%%%%%%%%%%%%%%%%%%%%%%%%%%%%%%%%%%%%%%%%%%%%%%%%%
\begin{abstract}
Lead pipe remediation budgets are limited and ought to maximize public health impact.
This goal implies a non-trivial optimization problem; lead service lines connect water mains to individual houses, but any realistic replacement strategy must batch replacements at a larger scale.
Additionally, planners typically lack a principled method for comparing the relative public health value of potential interventions and often plan projects based on non-health factors.
This paper describes a simple process for estimating child health impact at a parcel level by cleaning and synthesizing municipal datasets that are commonly available but seldom joined due to data quality issues.
Using geocoding as the core record linkage mechanism, parcel-level toxicity data can be combined with school enrollment records to indicate where young children and lead lines coexist.
A harm metric of estimated exposure-years is described at the parcel level, which can then be aggregated to the project level and minimized globally by posing project selection as a 0/1 knapsack problem.
Simplifying further for use by non-experts, the implied linear programming relaxation is solved intuitively with the greedy algorithm;
ordering projects by benefit cost ratio produces a priority list which planners can then consider holistically alongside harder to quantify factors.
A case study demonstrates the successful application of this framework to a small U.S.~city's existing data to prioritize federal infrastructure funding.
While this paper focuses on lead in drinking water, the approach readily generalizes to other sources of residential toxicity with disproportionate impact on children.
\end{abstract}


\maketitle
%%%%%%%%%%%%%%%%%%%%%%%%%%%%%%%%%%%%%%%%%%%%%%%%%%%%%%%%%%%%%%%%%%%%%%%%%%%%%

\section{Introduction}

\begin{figure}
\centering
  \includegraphics[width=0.75\linewidth]{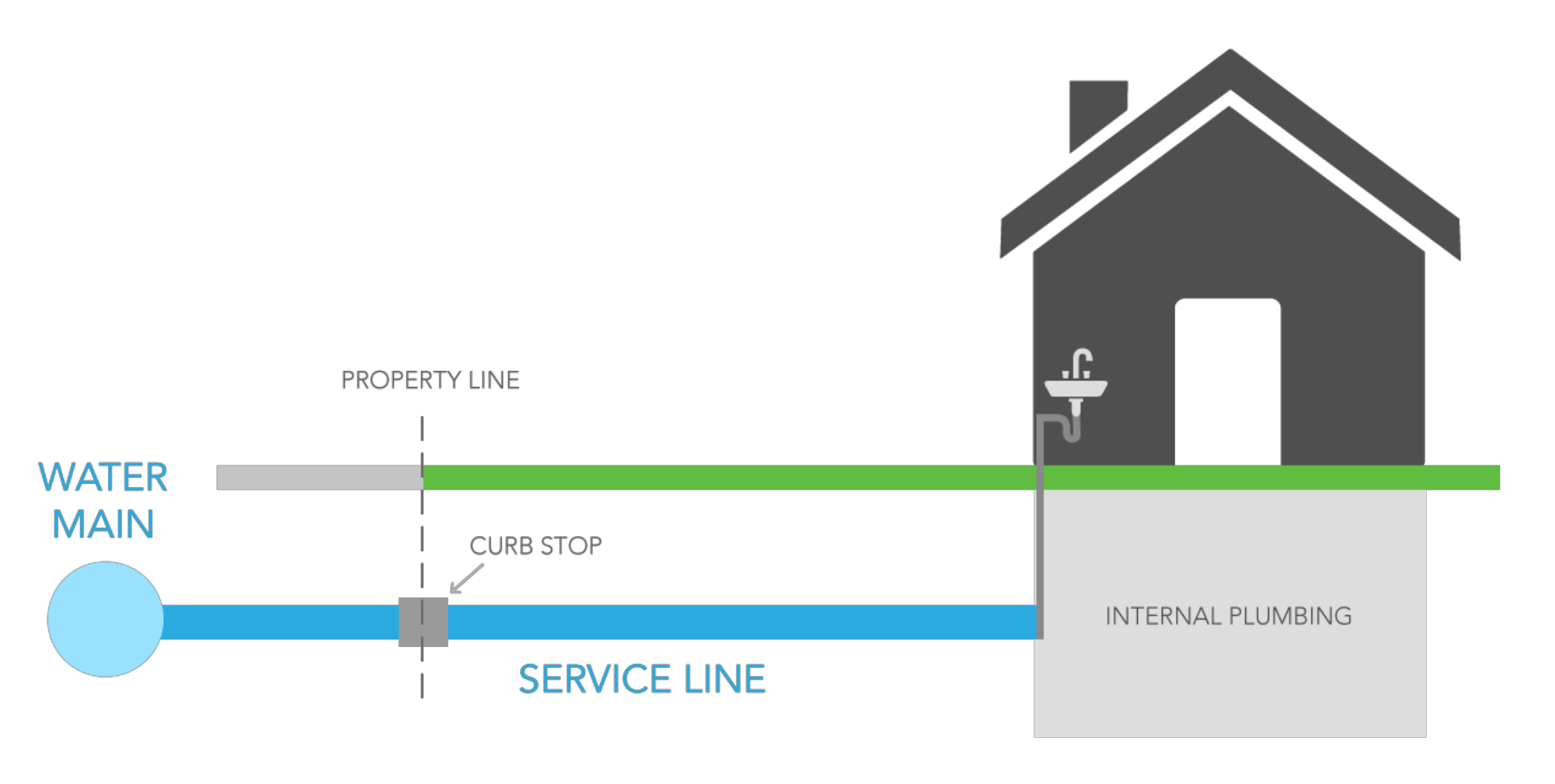}
\caption{Diagram of water service lines. Reproduced with permission from the \href{https://www.lslr-collaborative.org/intro-to-lsl-replacement.html}{LSLR Collaborative}.}.
\label{fig:lslr}
\end{figure}

Childhood exposure to lead is associated with catastrophic health outcomes,
particularly in early childhood, and one of the most common causes of lead toxicity among children is household drinking water service lines made of lead as illustrated in Figure \ref{fig:lslr} \citep{cdc_2012,zartarian_2017}.
Cities typically have a goal---or a legal requirement---to complete enough lead service line replacement (LSLR) projects to eventually replace all harmful lines.
But how should these multi-parcel projects be triaged?

At present, most planners lack both a conceptual framework and the clean data that would enable better replacement policies.
Even a simple heuristic such as ``prioritize streets with many children first'' is difficult to implement in the United States because there is no central database of individual demographic information accessible to a municipality, particularly for minor children who do not yet pay taxes, vote, or receive mail.
Without a convincing method for estimating impact, city staff tend to schedule remediation projects for non-health reasons such as alternating between neighborhoods, or by which streets already need repaving or utility work.
If they have reliable records for underground water infrastructure they may try to factor in the raw number of public and private lead lines which a given project could remove.
However, such a policy still misses the disproportionate public health impact of lead exposure on young children \citep{hu_2007,hhs_2020}.

Establishing a transparent and reproducible way to quantify and subsequently to maximize childhood harm reduction over time would equip city decision makers with impartial criteria for directing limited resources to help the most vulnerable residents.
It would also help them estimate the return on investment of any surplus funding, enabling data oriented advocacy for grant applications and budgeting processes.
Thus, this paper shows how geocoding can turn messy municipal datasets into a directionally correct proxy for where children live serving as the basis to prioritize project selection for maximal public health impact.

\paragraph{Primary goal: a flexible and data-driven way to estimate impact and prioritize projects}
This paper describes an exposure-based estimation approach by first assuming idealized data and demonstrating how the knapsack algorithm can conceptually be applied to maximize public health harm reduction subject to a limited budget.
The knapsack requires discrete projects with a known cost and value, so estimations for these are discussed including a novel estimation for project value based on toxicity exposure-years estimated using students' current grade in school.
Given the various estimation steps, the setup is accordingly relaxed from optimal binary assignments to a priority ordering such that planners can make holistic decisions considering multiple softer objectives which lack clean data or are too complex to model.
While this paper focuses on lead service lines, the framework could generalize to numerous toxicity mitigation efforts which present a similar modeling question, including remediation of residential lead paint as described in \citet{potash_2020}, or contamination from industrial, military, or resource extraction activity.

\paragraph{Secondary goal: making challenging city datasets useful for this purpose}
Though the proposed method for estimating impact is straightforward, the limiting factor in any real application will almost certainly be the availability and quality of extant data.
Like many organizations without a permanent cadre of database administrators and software engineers, municipalities tend to have heterogeneous data of uneven quality spread across numerous Excel spreadsheets, Access databases, PDF documents, GIS shapefiles, and proprietary formats.
Joining together their records for school enrollment (tabular data), residential parcels (geospatial data), and public works records on lead lines (various formats if available) presents a data cleaning challenge.
Section \ref{municipal-record-linkage} describes a simple record linkage approach using address normalization and geocoding to cross-reference typically noisy but commonly available municipal datasets in order to map out where children live.

\section{Methods}
\label{methods}

\subsection{Case study: the most affected city in Massachusetts}

The prioritization framework outlined here developed out of research in Malden, Massachusetts---a small city in the Boston metropolitan area with around 66,000 residents and ``the community [in the state] with the highest percentage of service lines made of lead'' \citep{globe_2016}.
Since 2006, Malden has been under an administrative consent order from the Massachusetts Department of Environmental Protection which requires the city to replace at least 150 lead service lines each year, but there are thousands of extant lines with known or likely lead status \citep{omalley_2017}.
To expedite replacements with the highest return on investment, the city is actively using this framework to prioritize over \$2 million allocated under the American Rescue Plan Act in 2022 with additional funds anticipated from pending legislation over the next several years \citep{gabrilska,advocate_2021,christenson_2022}.
Discussion of this case study will be interspersed with descriptions of methods in order to illustrate various practical considerations and implementation details.

\subsection{Modeling as a relaxed knapsack problem}
\label{knapsack}

Before describing how we will overcome specific data challenges, it will be helpful to sketch out how mitigation projects should be chosen in the ideal case.
Assume the set of all parcels with public or private lead service lines can be partitioned sensibly into $N$ mutually exclusive potential projects.
Each potential project $P_i$ constitutes the replacement of all public and private lead service lines in associated subset of parcels $i$ for $i \in \{1, \ldots, N\}$.
Selecting project $P_i$ incurs a financial cost of $c(P_i)$ and realizes a harm reduction value $v(P_i)$.
For now, this value function is left unitless and unknown, but later we define a heuristic $\hat v(P_i)$ for estimating the value of a project based on estimated future years of child exposure.

We can then specify an objective function and set up an integer linear program to pick projects maximizing total public health value subject to a fixed budget $W$.
Let $x_i \in \{0,1\}$ for all $i \in \{1, \ldots, N\}$ be binary decision variables denoting whether project $P_i$ will be completed. 
We want to maximize the total value by selecting the best subset of projects without going over budget:
\begin{align*}
  \mathrm{maximize} & \sum_{i=1}^{N} {v}(P_i) \cdot x_{i} \\
  \mathrm{subject\ to} & \sum_{i=1}^{N} c(P_i) \cdot x_i \leq W
\end{align*}

\noindent This sketch defines a classic 0/1 knapsack problem \citep{martello_knapsack_1990,kellerer_knapsack_2004}. However, while our projects are not continuously divisible and we do not intend to subdivide projects, a project $P_i$ can theoretically be subdivided down to the level of whether each individual parcel in subset $i$ is remediated.
That means that our application is closer to the \textit{linear programming relaxation}---also known as the continuous or fractional knapsack---where partial selections are allowed ``by removing the integrality constraint'' \citep[sec.~2.2]{kellerer_knapsack_2004}.
Though our goal is still to select entire projects, this fact will become important in section \ref{prioritizing} when we must select a solution approach.

In theory, this basic modeling setup could be extended until the problem was fully encoded in the constraints and objective, for example by discouraging road disruption in the objective function or adding a constraint to ensure the city is in compliance with the legally mandated annual minimum number of line replacements.
But choosing the strict optimization path requires being able to articulate all of the city's considerations in exacting detail and specify them with the right data, while in practice planners must simultaneously balance many ``soft'' objectives and constraints.
For example, engineers try to minimize disruption to residents; roads that have been recently resurfaced are less attractive to tear up, and roads in bad condition or with utility projects already scheduled may be more convenient to schedule for excavation concurrent with other work.
Factoring these sorts of secondary goals into a comprehensive project cost or encoding them as hard constraints would likely be impractical because of missing data or because they are too hard to distill algebraically.

Therefore, while grounding the approach in the well understood and conceptually appropriate knapsack constraints and objective, we propose a much simpler strategy of prioritization over strict optimization solutions.
Planners can take this ranking into account alongside all of the other factors they must consider, selecting projects until the budget is consumed.
Without loss of generality, there are four jobs to be done in any application of this prioritization framework:

\begin{enumerate}
    \item \textbf{Partition projects.} Divide the space into candidate projects $P_i$.
    \item \textbf{Estimate values.} Find a way to estimate project values $v(P_i)$ in terms of a heuristic $\hat v(P_i)$.
    \item \textbf{Estimate costs.} Find a way to estimate project cost $c(P_i)$ with a heuristic $\hat c(P_i)$ or ensure that the partitioning results in potential projects of approximately equivalent cost.
    \item \textbf{Prioritize projects.} Sort projects into a priority list decision makers can weigh alongside other considerations.
\end{enumerate}

\noindent While the project partitioning and cost estimation steps described here may vary with local conditions and available data, section \ref{exposure-length} establishes an extremely simple proxy for harm that can likely generalize to other residential risk modalities as a sensible default when no more specific way of calculating harm is available.

\begin{figure}
\centering
  \includegraphics[width=\textwidth]{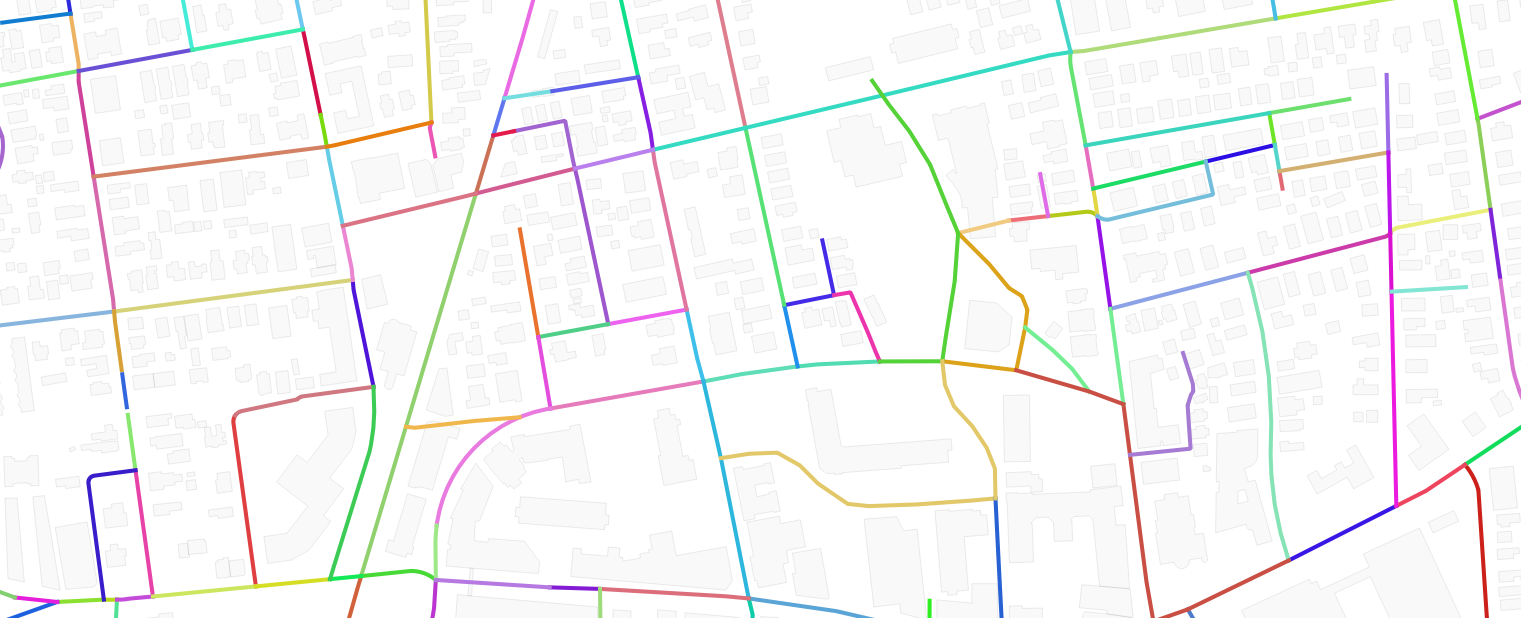}
\caption{Sample detail from the Malden ``centerlines'' shapefile. This segmentation was already used by the city engineering department for planning water projects. Random colors show how streets were segmented and a mix of residential and commercial building footprints indicate rough scale.}
\label{fig:street-segments}
\end{figure}

\subsection{Step 1: Partitioning the problem space into potential projects}
\label{partitioning}

Projects typically entail excavating a discrete section of road for replacement work at multiple parcels and then road resurfacing after replacements are complete, so a natural way to partition cities into mutually exclusive projects is to divide streets into contiguous segments.
The simplest division method, and one already familiar to planners, is to split streets into segments at intersections or water main junctions while limiting projects to reasonable sizes, which for LSLR projects tends to be about one city block (100 to 200 meters).

To then determine which parcels would be affected by each project, we need to know which parcels are physically adjacent.
For municipalities already using geographic information systems (GIS), these parcel maps are often maintained in electronic form with exact property footprints referenced to standard geographic datum points.
Where such GIS data does not exist, the same approach to geocoding described in section \ref{municipal-record-linkage} can be used to place a centroid point on the map for each parcel, for which the closest perpendicular point on the street segment matching the mailing address can be found.\
More advanced approaches might attempt to place project endpoints dynamically such that each project has a similar number of lines, or even allow project partitioning to be parameterized and simultaneously optimized with selection; these possible extensions are discussed in section \ref{future-work}.

In Malden, the GIS office already had a shapefile of street segments divided for water infrastructure engineering purposes, which for operational reasons made the most sense to adopt in this application. Had that not been available, it would have been straightforward to take an open dataset such as \href{https://www.openstreetmap.org/}{OpenStreetMap} and split streets into smaller segments, for example wherever cut by intersections or upon reaching a maximum straight length. Similarly, the U.S.~Census Topologically Integrated Geographic Encoding and Referencing (TIGER) dataset provides shapefiles of road segments terminating in intersections which could be useful for this purpose \citep{census_tiger}.

\subsection{Step 2: Estimating project value with a simple proxy for childhood exposure}
\label{project-value}

Given perfect information, an idealized formula for project value could attempt to capture each associated parcel's exact lead prevalence, the physical conditions of piping and joints, and the resulting interaction with each affected person's age, biological susceptibility, personal habits, length of exposure, and so forth. These factors would all be inputs to a function quantifying the instantaneous harm to any individual at any particular moment.
In reality almost none of these quantities of interest will be known, and an instantaneous harm function is neither feasible to compute nor an accurate representation of the uncertainties in the way lead particles interact with the body over prolonged periods of time \citep{hu_2007,hhs_2020,lockitch_1993}.

Any practical approach must establish a simplified metric using data likely to be available in the field.
One simple proxy suggested by the toxicology literature for the amount of harm caused by a slowly accreting toxin is length of exposure, which for the purposes of this application is further narrowed to the length of exposure to children \citep{whitmyre,cdc_2012}.
When combined with information about where lead lines might be, if available, length of exposure presents a sensible stand-in for actual health impact.

\subsubsection{Estimated length of exposure as a heuristic for project value}
\label{exposure-length}

Each planning cycle, analysts can obtain a fresh snapshot of which children are currently being exposed to lead service lines.
For each child exposed, we can add the number of years $T_j$ over which the child would be exposed if the mitigation project is not pursued.
This results in a simple estimation for the value of project $P_i$: removing $T_j$ years of exposure for each of $j \in C_i$ children living at any of the parcels which project $P_i$ addresses \textit{if} the child is also in the set $L$ of children living at parcels with a lead service line:
$$
  \hat v(P_i) \coloneqq \sum_{j \in C_i \cap L} T_j
$$

\noindent The next two sections discuss prerequisites for finding the set $L$, but first we must find a way to sidestep the limitation that actual future exposure duration is unknown. We do know each child's age $a_j$ and can introduce an assumption that a child will leave the childhood home around the age of eighteen, which is both the last age year at which high school data will typically be available and close to the prevailing median value in the United States \citep{census_cps}. In any case, this curtailed estimate will serve the purpose of targeting work towards young children:
$$ T_j \approx%
  \begin{cases}
    18 - a_j & a_j < 18 \\
    0 & a_j \geq 18
  \end{cases}
$$

\noindent If age is not known but school grade is---typical of anonymized data without birth dates that schools are willing to release---$a_j$ may be estimated by using the expected age in grade.
Further, if the city knows that, for example, all lead lines will have been replaced in ten years, then the years of exposure can be capped at this maximum duration.

\subsubsection{Finding which parcels have lead lines}
\label{finding-lead-status}

Many municipalities with ongoing lead remediation efforts already maintain an inventory of lead status by parcel, and recent federal regulations in the United States require municipalities to have an inventory of lead pipes by October 2024 \citep{epa_lcrr}.
In the meantime, if a city has no inventory of lead lines or if the inventory has many missing values, a number of decision policies with different tradeoffs can be applied to lines with unknown status.
Even if planners start out lacking any reliable data about which services lines are lead, they can still use the child location data described in the next section to prioritize water quality testing or data gathering excavations.

Heuristic policies for unknown lead status include assuming lead for parcels built before a certain date, assigning an intermediate value to unknown lines in order to gently encourage selection, or more conservatively treating all unknown lines as if they were lead.
More advanced forecasting methods have also proven to be effective for this purpose; there is a growing body of research on machine learning approaches to predicting which parcels are most likely to have lead lines given year built, proximity to other properties with known lead, and numerous other relevant features \citep{abernethy,goovaerts_2017}.
If model outputs are probabilities or have meaningful relative size, each parcel's contribution to the corresponding $\hat v(P_i)$ can be scaled by its likelihood of having lead service lines.

The City of Malden already had a list of public and private side lead line status by parcel, albeit with a significant number listed as unknown. Table \ref{table:counts} shows a recent snapshot of status counts broken down by public and private side.
The city also hosts an ArcGIS web application with this data displayed on a map so that residents can look up the status of their home, and another GIS web application that serves a clean inventory of Malden property parcels.
Data from these applications were straightforward to obtain.

\begin{table}
\centering
\caption{Counts of service lines by material and public/private side in Malden as of September 2021.}
\label{table:counts}
\footnotesize
\begin{tabular}{lrrrrrrrrr}
\toprule
\backslashbox{\textbf{Private}}{\textbf{City}} &  brass &  iron &  copper &  lead &  PVC &  steel &  unknown \\
\midrule
          iron &      0 &         46 &      77 &     5 &    0 &      2 &       87 \\
        copper &      5 &          6 &    5,516 &    975 &    2 &     24 &     1,275 \\
         steel &      0 &          0 &       0 &     0 &    0 &      0 &        3 \\
          lead &      1 &          0 &    1,415 &   267 &    0 &      6 &     1,049 \\
       unknown &      0 &          1 &      32 &     7 &    0 &      0 &      282 \\
\bottomrule
\end{tabular}
\end{table}

\subsubsection{Finding which parcels have children present}

School districts must maintain lists of enrolled students for legal and correspondence purposes, and these lists necessarily include the student's school grade (a proxy for age if date of birth is not available) and parents' mailing address (which is usually the residential address unless tracked separately). School rolls are updated at least annually, and given that families move around and children continue to age, the outputs of this analysis will vary from year to year and should be kept updated while being used to inform operational decisions.

In Malden, the Superintendent of Schools maintains a master list of all public school students. This data includes a number of fields that are considered sensitive and legally protected, but the simplicity of the proposed estimation method means that it does not require any personal information except school year and street address, so school administrators were able to share an anonymized extract for research purposes. The primary challenge is in linking this student data to parcels, which is discussed at length in section \ref{municipal-record-linkage}.

\subsection{Step 3: Estimating project cost}
\label{project-cost}

Accurately assessing the predicted cost of a potential project is an expensive and time consuming undertaking requiring engineers or contractors to visit the site, evaluate on-scene conditions, and possibly conduct a more thorough investigation into city records or solicit competitive bids from contractors.
Exhaustively capturing all anticipated costs $c(P_i)$ across the city would not generally be feasible, so we instead substitute a heuristic $\hat c(P_i)$ as a proxy for the estimated cost of a project.

Discussions with city engineers elicited that project costs can be estimated using $d$ dollars per service line (private or public) being replaced plus a fixed overhead cost $K$ for any project regardless of size: $$ \hat c(P_i) \coloneqq d \cdot \lvert P_i \rvert + K $$

\noindent These costs will vary depending on the specifics of engineering and contracting practices, and can be adjusted for local conditions.
Although the fixed cost $K$ would tend to discourage selection of the smallest projects under the full optimization setup, such costs will be small compared to variable costs for any non-trivial project and are therefore less important to capture.
If neglecting fixed costs, the constant coefficient $d$ will not matter for relative comparisons so that the total number of lead service lines per segment can serve as a useful first approximation for project cost: $$ \hat c(P_i) \approx \lvert P_i \rvert $$

\noindent This is about as simple a cost estimate as possible, but still captures projects' relative order of magnitude and as a side benefit is useful to planners as a proxy for the amount of disruption caused to a particular street. Given the data we have already assembled in previous sections, planners could choose to factor in more granular costs if desired.
If lead line status is not known or if missing values are problematic, the total linear street length of the project can be used as a similar but even simpler proxy for cost.

\subsection{Step 4: Prioritizing potential projects by benefit cost ratio}
\label{prioritizing}

Although it is possible to solve the 0/1 knapsack posed in section \ref{knapsack} optimally, lack of strict optimality is not a concern in this application given the limited input precision from noise in the data and multiple estimation steps;
as observed by \citet{sahni_1975}, ``it is often the case that obtaining exact solutions to the knapsack problem may not be necessary [...] for instance, when the profits, $p_i$, themselves are only estimates of expected returns or when the knapsack problem itself is only a suboptimization of a much larger problem.''
Planners must also balance a variety of unmodeled goals, constraints, and unpredictable logistical developments simultaneously, so a flexible ordering will be more useful than recommending an exact, unordered set of projects that must be completed in the current time period.

For these reasons, we prefer an approximate solution that will result in a priority list of all potential projects.
Many approximation algorithms rely on sorting candidates by the \textit{benefit cost ratio}, which \citet{dantzig_discrete-variable_1957} originally called the \textit{effective gradient} but is also variously called the \textit{bang for buck}, \textit{efficiency}, or \textit{profit ratio} \citep[see, for example,][]{senju_approach_1968,kellerer_knapsack_2004,martello_knapsack_1990}:

$$\frac{\hat v(P_{1^*})}{\hat c(P_{1^*})} > \frac{\hat v(P_{2^*})}{\hat c(P_{2^*})} > \ldots \frac{\hat v(P_{N^*})}{\hat c(P_{N^*})}$$

Once these ratios are calculated and sorted, the simplest approach to project selection is the \textit{greedy algorithm}:
``If a non-expert were trying to ﬁnd a good solution for the knapsack problem [...] an intuitive approach would be to consider the proﬁt to weight ratio [...] and try to put the items with highest efﬁciency into the knapsack. Clearly, these items generate the highest proﬁt while consuming the lowest amount of capacity.'' \citep[sec.~2.1]{kellerer_knapsack_2004}
While the greedy algorithm is not generally an optimal solution to the 0/1 knapsack problem \citep[sec.~2.4]{martello_knapsack_1990}, it is still an effective choice for this application for several reasons.

First, sorting potential projects by benefit cost ratio is supported by a body of literature in ecological conservation and environmental management \citep[see][]{joseph_optimal_2009,pannell_ranking_2015,beher_prioritising_2016} and engineering \citep{koc_prioritizing_2009}.
Also, as noted in section \ref{knapsack} this application is closer to the continuous knapsack than the 0/1 formulation, and the greedy algorithm applied to the continuous linear programming relaxation is provably optimal as long as the last selected project may be subdivided \citep[sec.~2.2]{kellerer_knapsack_2004}. Even if planners do not actually intend to subdivide the last project, empirical research into similar applications by \citet{pannell_gibson_2016} has shown that greedy prioritization on the 0/1 knapsack tends to display negligible underperformance despite theoretical worst case bounds, so the difference will likely be well within the margin of estimation error and thus not a motivation for more complex approaches.

Perhaps most importantly, ordering by ``bang for buck'' is intuitive and easily explained to non-expert stakeholders in terms of return on investment, a crucial aspect captured well by the adage that ``a decision maker would rather live with a problem he cannot solve than a solution he cannot understand'' \citep{eiselt_multicriteria_2014}. Given that results in the motivating application would be subject to a potentially adversarial budgeting and deliberative process, analysts and decision makers must be able to explain and defend outputs without reference to the theory of combinatorial optimization.

\section{Municipal record linkage through geocoding}
\label{municipal-record-linkage}

The primary difficulty in implementing the estimates described above is that the necessary datasets are ordinarily intended for narrow administrative uses and are therefore difficult to link.
In theory, each student record being added or updated in the school enrollment database could be enriched with a parcel number uniquely identifying the student's primary physical residence.
In practice, school administrators do not track this because they do not need to know the student's geographic parcel once it has been established that they are eligible for enrollment in the district and that letters can be mailed to parents.

Even though the datasets needed to apply this methodology were not configured with cross-domain analysis in mind, they each generally contain street addresses which serve as a starting point for reconciliation.
Students live in particular buildings, these buildings have water service lines, and the municipality keeps records on service lines.
The common thread between each relevant dataset is the \textit{residence}, and every ordinary residence has a street address.
However, because record linkage is constrained by the lowest quality data the address information in original form will not usually be sufficient for automatic linkage without further cleaning and enrichment.

Analysts must be prepared for the extremely common case where the address is an unstructured or semi-structured text string, neither coded to a particular latitude and longitude nor indexed to a particular parcel of land in the assessor's database.
Even when datasets break addresses into structured fields such as street 1, street 2, city, postal code, etc, such divisions are error prone and inconsistent, and the most important part of the address if operating within one city---the street address---is still free entry text.
Any amount of free entry text poses a thorny data quality challenge given the frequent defects it entails in real data: simple typos (``Peasant St'' versus ``Pleasant St'');
variations in street number, both meaningful (``3'' versus ``3R'') and exchangeable (``\#3'' versus ``Apt 3'');
variations in suffix, both trivial (``Street'' versus ``St'') and meaningful (``St'' versus ``Rd'', ``St'' versus ``St Ext,'' especially problematic where both variations may exist on the map).
It is even common to observe profound errors such as incorrect town or postal code.

Geocoding and normalization provide a way to automatically link the majority of records so that they can be combined for analysis.
There exist a number of services and open source packages which take arbitrary text and attempt to recommend the most probable canonical locations, often having quantified the likelihood of each candidate match \citep{goldberg_2017}.
Geocoding services attempt to correct all of the errors described above through \textit{normalization} where common variations are standardized (e.g., ``St.'' and ``Street'' both mapping to \texttt{ST}) and \textit{inference} where impossible or improbable parts of the text are replaced with their most likely intended value.

We can refer to the output of geocoding as a \texttt{GeoAddress} data record. The content of this \texttt{GeoAddress} varies between geocoding services, but for the purposes of discussion it can be considered a highly structured data record with a globally unique identifier, a geospatial point or polygon, and text address information broken down into rigid fields for city, postal code, country, and intermediate administrative divisions like state and county.
An example response payload in \nameref{appendix:payload} shows the result of geocoding a free entry text string with an intentional typo.

While implementations differ, this transformation can be represented abstractly as a mapping from arbitrary strings of text to a set of structured records each representing a single address:
$$\text{geocode}( \cdot ) \colon \texttt{Text} \mapsto \begin{cases}
    \texttt{Set} \left [ \texttt{GeoAddress} \times P(\text{match}) \right ] & \text{if any matches found} \\
    \emptyset & \text{otherwise}
 \end{cases} $$

\noindent Crucially, this maps the set of all possible text strings into the clean set of geospatial places of which the geocoding service is aware, meaning that multiple variations of the same address can be \textit{canonicalized} to a single representation.

\begin{figure}
\centering
  \includegraphics[width=\textwidth]{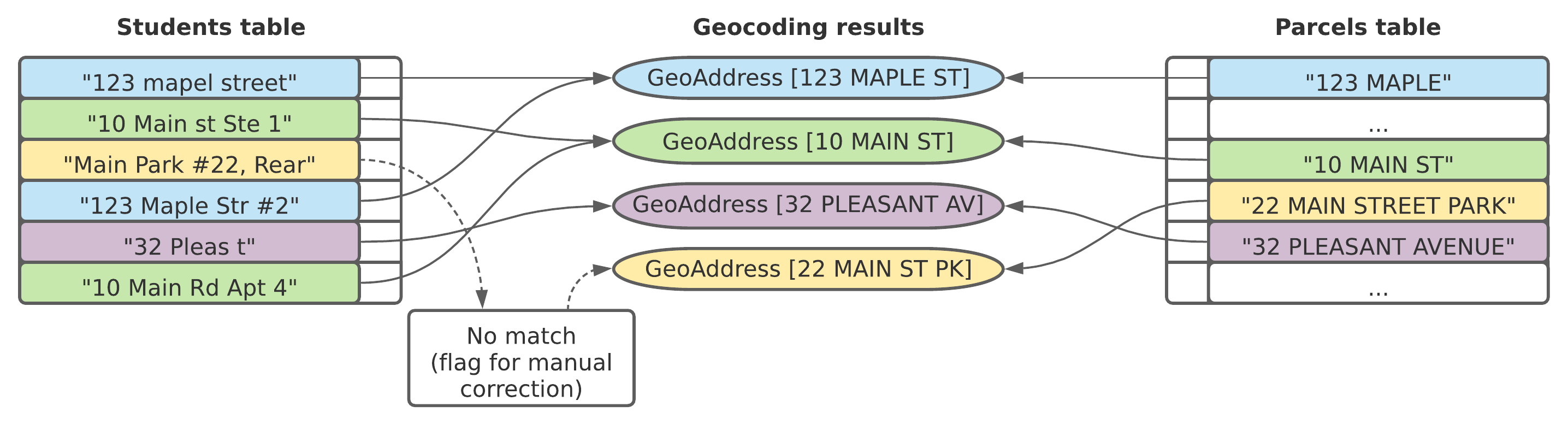}
\caption{Geocoding all addresses in tables with text addresses allows the mapping of results to serve as a \textit{junction table} (also known as a \textit{joining table}) between the two original tables. The fraction of addresses which fail to geocode properly can be quickly fixed with manual correction.}
\label{fig:geocoding}
\end{figure}

An output payload can be requested for each address in all of the datasets that we need to link, and once these complete geocoding results have been associated with every original record we can ignore the free entry text addresses and treat either the single most likely or the top $k$ most likely \texttt{GeoAddress} records as joining keys as illustrated in Figure \ref{fig:geocoding}.
The minority of records which cannot be automatically linked may be corrected by hand if desired or omitted if exactness is not necessary.
Between mostly automatic matching and some manual correction, this reliable record linkage mechanism unites the datasets necessary for this framework and---by referencing the record to a geospatial datum---many more datasets that may be helpful in extending this work.

\section{Results and impact}
\label{results}

While it makes intuitive sense that focusing on streets with the youngest children is a better way to target projects than the status quo, it is still useful to characterize how much of an improvement this methodology enables.
Using data from the case study, we can compare the consequences of various replacement policies in terms of the definition of impact proposed above.
Even though the data is incomplete (e.g., infants, children who attend private or parochial schools that did not voluntarily contribute data, etc), this will still provide a directional indicator of impact to the children we do know about, as well as illustrating the distribution of impact among potential projects.
Additionally, section \ref{future-work} describes how this work may be extended to cover more affected individuals and increase the fidelity of estimated quantities.

\subsection{The long tail of project value}

\begin{figure}
\centering
    \begin{minipage}{0.66\textwidth}
        \centering
        \includegraphics[width=\textwidth]{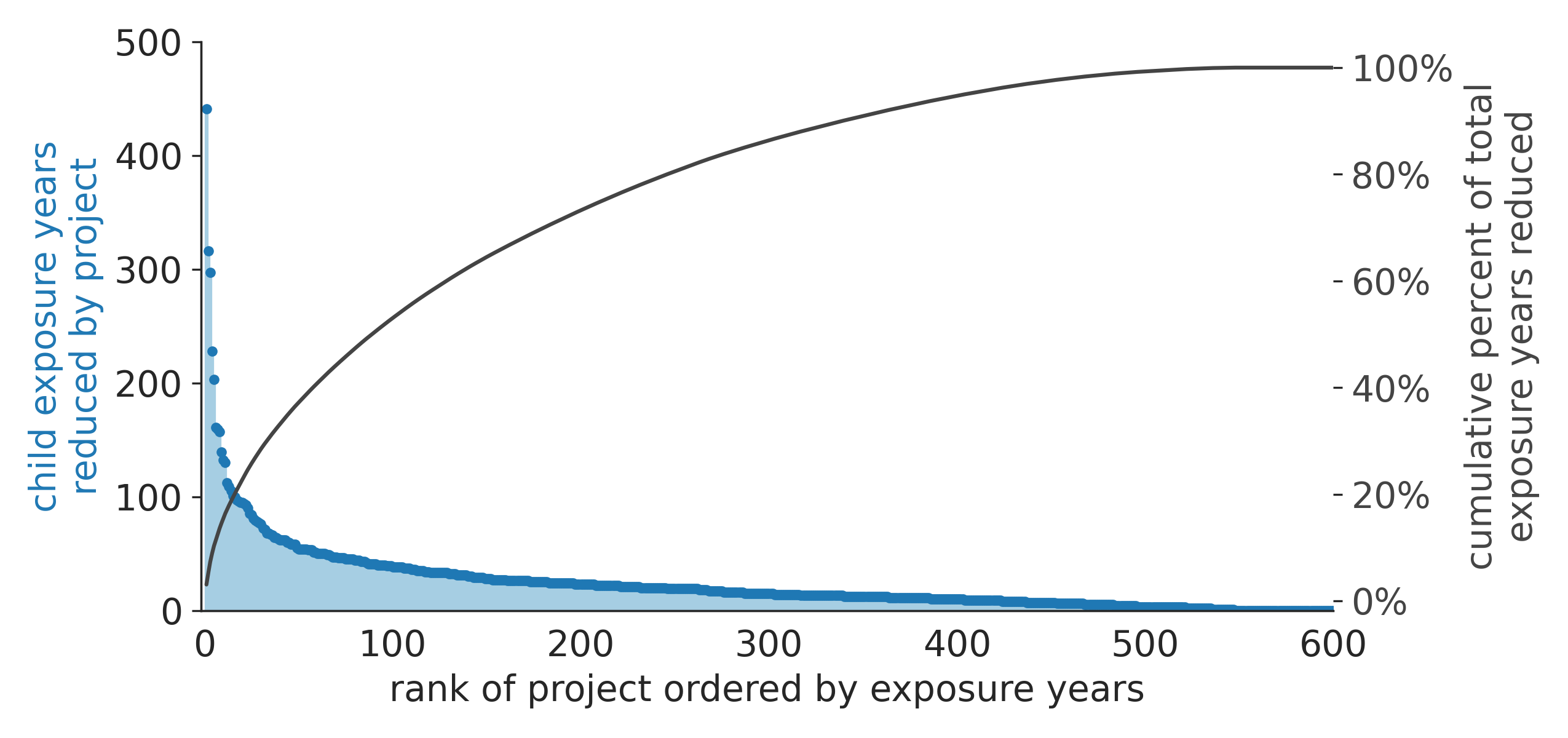}
    \end{minipage}\hfill
    \begin{minipage}{0.33\textwidth}
        \centering
            \footnotesize
            \begin{tabular}{rr}
                $n^{\text{th}}$ project & cumulative \% \\
                \midrule
                5 & 10.3\% \\
                10 & 15.5\% \\
                20 & 22.7\% \\
                50 & 37.2\% \\
                100 & 53.0\% \\
                150 & 64.5\% \\
                200 & 73.2\% \\
                300 & 86.4\% \\
                400 & 94.7\% \\
                %\bottomrule
            \end{tabular}
            \vspace{16pt}
    \end{minipage}
    \caption{Estimated project impacts in Malden, Massachusetts displaying a power law distribution when ordered by total exposure years. The figure shows both individual project exposure year impacts (in blue, left vertical axis) and cumulative percent of total exposure years represented (in black, right vertical axis) by the project's order in the priority list (horizontal axis). The table on the right shows selected point values from the figure.}
    \label{fig:cumulative-exposure}
\end{figure}

The first insight provided by the cleaned data is that distribution of project value is far from uniform. In fact, as Figure \ref{fig:cumulative-exposure} shows, plotting project impacts in descending order reveals a classic power law distribution, with a small number of highly impactful projects and a ``long tail'' of projects diminishing to almost negligible impact.
As the figure shows, the first 25 projects account for the first quartile of exposure years in the entire dataset; the next 65 projects account for the second quartile of exposure years; after that it takes an additional 120 projects to get the third quartile of exposure years, while the last quartile of exposure years diminish increasingly slowly with the completion of hundreds more projects. The table in Figure \ref{fig:cumulative-exposure} provides more detail on this long tail in action.
Even though our dataset is incomplete and relies on imperfect estimates, the figure illustrates that some projects present dramatically higher return on investment, and should persuade stakeholders of the general principle that it is worthwhile to prioritize some projects over others for targeted health impacts.

\subsection{Simulating the consequences of different replacement policies}

\begin{figure}
    \centering
    \begin{minipage}{0.66\textwidth}
        \centering
        \includegraphics[width=\textwidth]{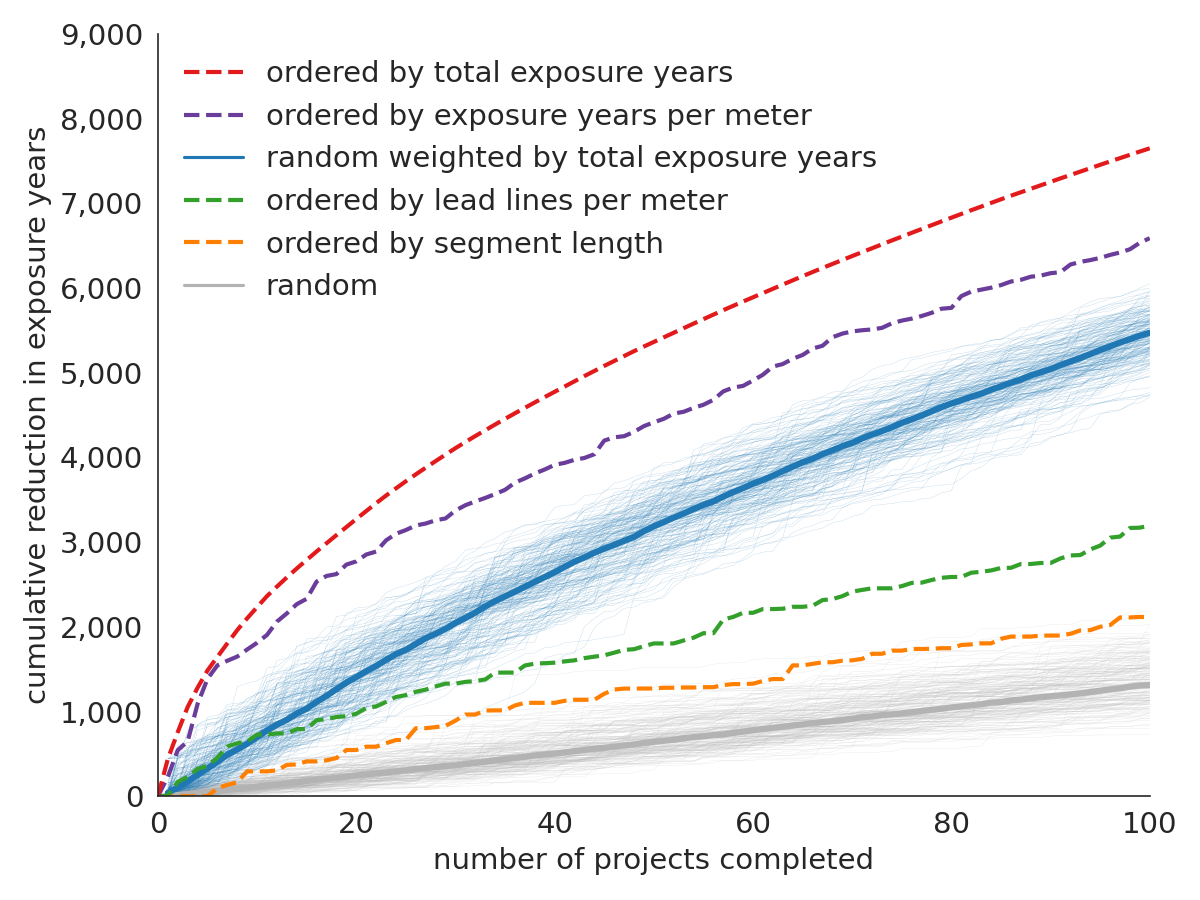}
    \end{minipage}\hfill
    \begin{minipage}{0.33\textwidth}
        \centering
        \includegraphics[width=\textwidth]{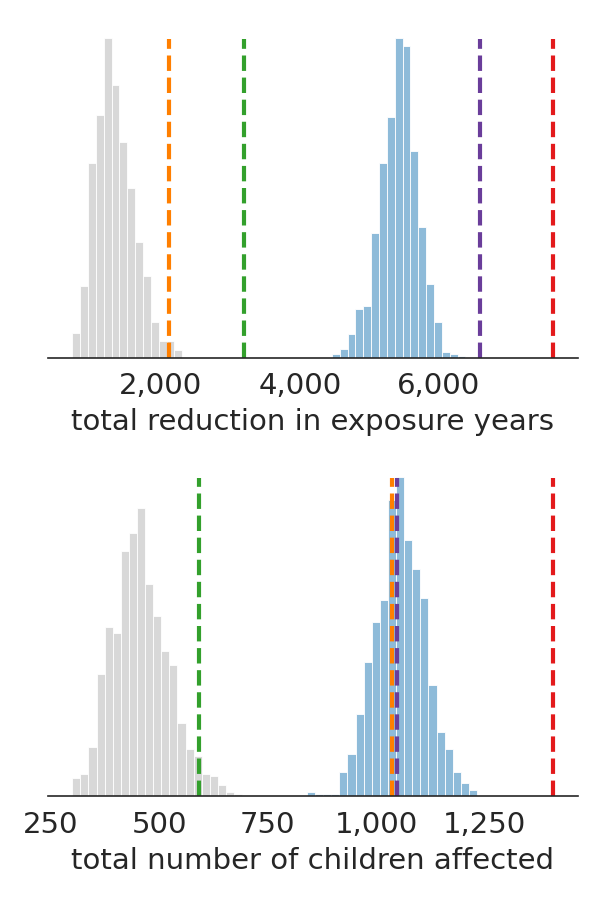}
    \end{minipage}
    \caption{Simulated impact of 100 projects on child exposure years in Malden, Massachusetts under different project selection policies. Dashed lines in the figure on the left show the deterministic results of conducting replacement projects in strict order of prioritization by a certain quantity. Solid lines in the figure on the left (blue and gray) show the results of simulation, with thin lines showing individual iterations of the simulation and thick lines aggregating all the iterations by the median. Histograms on the right show the impact of these same policies on several quantities of interest after all 100 projects are complete.}
    \label{fig:simulation}
\end{figure}

Using the proposed methodology for estimating project value, we may want to compare replacement prioritization policies to see how quickly they ``buy down'' total public health risk throughout the city, all other factors being held equal. Using the cleaned data from the Malden case study, we can pick metrics of interest such as total reduction in exposure years or number of children affected and compare the results given a limited budget and the order in which each policy would assign projects to be completed. Figure \ref{fig:simulation} shows the result of carrying out projects in strict orderings (dashed lines) or allowing some probabilistic variation by simulation (solid lines).

In the figure, the gray solid lines show what happens when projects are chosen completely at random; this simulation was carried out by randomly selecting orderings of 100 projects with equal probabilities and without replacement. For the random policy, as with the other simulation, thin lines show results from individual simulation iterations to illustrate variation while the thick line shows the median across iterations. This policy sets a minimum baseline below which it would be hard to do worse without trying.

The next two lines above the random policy show what is possible while still remaining blind to where children live. The orange dashed line shows what happens when naively ordering by the size of the project in terms of length of street excavated, while the green dashed line presumes knowledge of where specific lead lines are and orders projects by number of lead lines remediated per meter excavated if the project were carried out.

The blue solid lines show simulated orderings which are \textit{informed by exposure years} but do not rigidly adhere to that quantity for strict ordering; the way this is simulated is that orderings of 100 projects are selected without replacement randomly but with selection probability weighted in proportion to the exposure years represented by that project. This may be considered a realistic representation of what is possible in practice, where planners take into account the ordering produced by this methodology but also consider numerous other factors external to the model.

Finally, the purple and red lines show idealized results of ordering by total exposure years reduced if the project is carried out, either taking cost into account (purple) or not (red). Specifically, the purple line shows the impact of one variation of the benefit cost ratio ordering recommended in section \ref{prioritizing}, while the red line demarcates the upper bound on impact if cost is disregarded and projects are sorted by value alone.

The age-informed policies strictly dominate the others which is not surprising given that they order projects by the quantity we are using as a metric for success.
More surprising is the significant gap between the best naive policies---those possible without knowing where children live---and even the worst iterations of the merely risk-informed policy represented by the blue lines.
In cities at all comparable to the case study, this comparison offers a strong rebuttal to the potential objection that it would be unnecessary to find where children actually live because they would be more or less uniformly distributed throughout the city and with relatively even proportions of lead line exposure.

\subsection{User interface for transparency and communication}

The outputs of this framework were used by planners in a way that affects public health outcomes and the obligation of public funds, so it was important to explain the methodology clearly and approachably without unnecessary reference to equations or technical jargon.
As noted by \citet{costa_applying_2014}, end users ``do not need to know the methods of mathematical modeling.
They interact with a software application that encapsulates the techniques of operations research and allows the processing and analysis of the data involved in decision making. [...] The quality of a decision support system is measured not only by the results it provides, but also by the ease of interaction among users, application, and data.''
Indeed, the literature on human-computer interfaces (HCI) in optimization decision support suggests that methods are less likely to be adopted without an approachable display of quantitative information. This common finding is summarized well with the observation that ``many heuristic and optimization-oriented systems that have been implemented for this type of [operational planning] problem have failed in part due to a poor graphical user interface'' \citep{maturana_vehicle_1995}.

\begin{figure}
\centering
  \includegraphics[width=\linewidth]{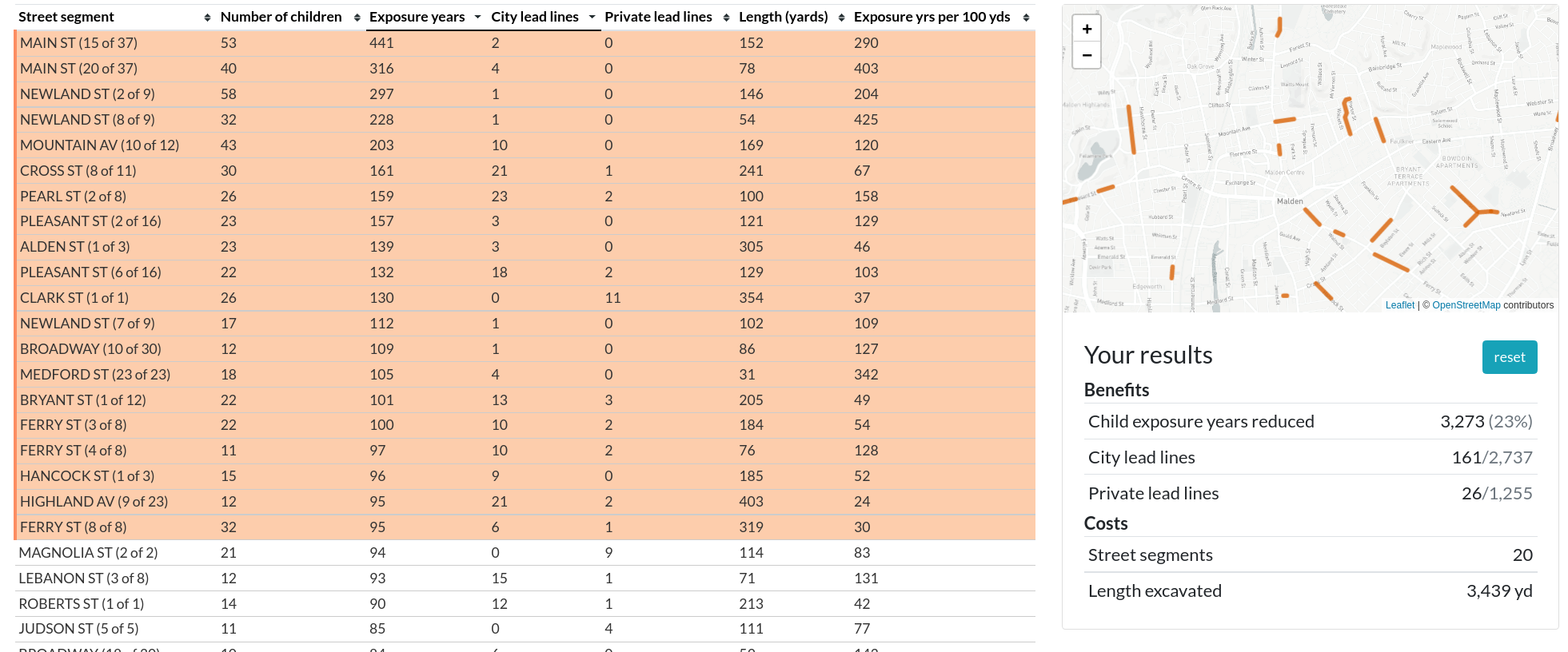}
\caption{Screenshot of the explorer tool. The website set up for publicly communicating this project at \href{https://www.maldenleadlines.org}{maldenleadlines.org} described the methodology and allowed users to experiment with different replacement choices.}
\label{fig:explorer}
\end{figure}

Accordingly, a web application was developed to communicate the methodology both in plain writing and through an interactive explorer tool which displays all of the potential projects in the city and allows a user to experiment with ordering projects in a variety of ways.
Users of the explorer tool can add street segments to a ``shopping cart'' of potential projects and observe the total number of exposure years which could be bought down by carrying out those projects.
Figure \ref{fig:explorer} shows the intuitive user interface with the only choice made by the user being whether to add a project to the worklist or not.
This choice interaction precisely mirrors the decision variables in the knapsack algorithm, making it an ideal analogy to build intuition about the problem setup.

Users reported that having a responsive sandbox in which to explore various replacement policies helped them think about the task as a city planner must, and helped center an uncontroversial principle (i.e.~that any replacement policy should safeguard children especially) as an objective starting point for prioritizing scarce public resources.

\section{Future work}
\label{future-work}

Many extensions to this work are possible which would address various assumptions and limitations described above. In particular, any estimation step would be well served by focused research to layer in additional fidelity. That said, there are several specific research directions which seem especially compelling.

\paragraph{Simultaneously optimizing project partitioning}
The way the solution space is divided into potential projects has a large impact on the upper bound of performance in the optimization step. As mentioned in section \ref{partitioning}, the partitioning of projects could be treated as a set of decision variables subject to optimization.
Instead of taking a preexisting street segmentation or cutting streets at intersections, the number and placement of cut points could be added as decision variables and optimized using a solver either simultaneously with the 0/1 knapsack or as a specialized optimization step.
For example, if there were two adjacent street segments where one had many lead lines and the other only had one located close to their mutual edge, the solver might find it worthwhile to expand or translate the first project endpoints to include that extra service line, thereby eliminating the smaller project.
Even if simultaneous optimization is not feasible, an initial step that induces more compact project packing would likely enable a faster remediation schedule.

\paragraph{Representing the challenges of private side replacements}
In the analysis presented it is assumed that private sides can be replaced at will, but in practice it can be difficult to persuade owners to pay for replacement, or to obtain consent for work on private property even if the replacement will be paid for with public funds.
Without that consent, the city should not replace the city side lines for which the private side is still lead because a \textit{partial replacement}---replacing only the public or the private side when both sides are lead---risks a net increase in water lead levels \citep{trueman_2016}.
Accordingly, the project cost estimate could factor in the overhead of obtaining consent for private side lines and the project value estimate could be tweaked so that private replacements are assigned a probabilistic value based on the likelihood of a successful replacement.
Recent work by Clean Water Action indicates that offering to pay for replacement of the private side expedites the process of getting owners to replace private lines, so extending the methodology to allow a comparison of policies could inform the public health case for using public funds to accelerate private replacements \citep{clean_water_action_2019}.

\paragraph{Quantifying service lines of unknown lead status}
As discussed in section \ref{finding-lead-status}, the methodology presented here assumes at least imperfect knowledge of where lead lines are on a parcel-by-parcel basis.
It would be helpful to extend the framework to situations where planners do not know which parcels have lead lines, as was the case in the Flint, Michigan crisis where research by \citet{abernethy} confirmed that lead status could be predicted with remarkable accuracy.
A natural extension would therefore be the ``predict then optimize'' strategy of adding a setup step where the presence of lead lines is estimated with statistical modeling as a stand-in or complement to existing data.
This would help address both the usual case where some number of lines are unknown as well the ``cold start'' problem of working in a city where lead status data is still sparse or low quality.

\paragraph{Differentiating the severity of exposure years at different ages}
How much worse is a given unit of lead exposure at age five as opposed to age seventeen? Currently, both years of exposure count as equally severe.
Finding a more accurate measure of exposure severity at different ages could re-weight exposure years in proportion to potential harm.
It is possible to pick an arbitrary decay curve instead of the linear assumption adopted in section \ref{project-value}, but basing the weights on a thorough review of the medical literature, if available, would enable a more refined estimate. The same research could be extended into adulthood so that the methodology would not be limited to children.

\paragraph{Developing more accurate estimates of where children live, including infants}
Even assuming constant severity by age, infants would have an outsize impact on exposure years if their locations were known or could be estimated.
Using the estimation described in section \ref{exposure-length}, a newborn at a parcel with lead lines would contribute eighteen full exposure years, an order of magnitude more than a high school student.
Although data may be difficult to obtain, it is possible that census, social security, or birth certificate records could be layered in to give a better picture of where the youngest children already live or are likely to be born.
Similarly, we could attempt to attribute exposure and thereby induce higher prioritization for locations where very young children are likely to spend time such as in-home daycare facilities.

\vspace{12pt}

\noindent Given how much of this framework involves estimating unknown quantities, it should be extremely easy for local applications to experiment with alternative methods---e.g., for locating children, partitioning projects, estimating cost and value---without changing the general approach, making each of these research areas independently and immediately useful.

\section{Conclusion}

Municipalities with lead service lines owe it to their most vulnerable residents to use all available data to inform sound public health decisions.
City planners want to be informed, but data quality and availability are typically the limiting factor on better policies.
In addition to providing a method to clean and link data, this paper provides a logical and simple framework for targeting limited resources where they will have the most impact.
The proposed approach intentionally trades away some exactness in the problem specification in favor of greater flexibility and usefulness in the field, but is highly extensible in municipalities which have more data or more ability to delve further into the technical implementation.
Even if project prioritization is approached in very different ways from the manner described in the case study, the main ideas of this methodology can help cities bring a fresh perspective to many residential health threats where children are most at risk.

\begin{acks}
This project would not have been possible without the enthusiastic cooperation of the City of Malden, particularly Mayor Gary Christenson and Maria Luise for support getting this project started and school data shared, Steve Fama in the GIS office for providing many datasets and insight into city recordkeeping, and the city engineer Yem Lip for his insights into LSLR project considerations.
Malden City Councillor Steve Winslow has been an advisor and advocate for this research as well as a steadfast voice for infrastructure modernization in the City of Malden.
\end{acks}

\bibliographystyle{itor}
\bibliography{article}

\begin{thebibliography}{33}
\expandafter\ifx\csname natexlab\endcsname\relax\def\natexlab#1{#1}\fi
\expandafter\ifx\csname url\endcsname\relax
  \def\url#1{\texttt{#1}}\fi
\expandafter\ifx\csname urlprefix\endcsname\relax\def\urlprefix{URL }\fi
\providecommand{\eprint}[2][]{\url{#2}}
%Type = Inproceedings
\bibitem[{Abernethy et~al.(2018)Abernethy, Chojnacki, Farahi, Schwartz and
  Webb}]{abernethy}
Abernethy, J., Chojnacki, A., Farahi, A., Schwartz, E., Webb, J., 2018.
\newblock {ActiveRemediation}: The search for lead pipes in {Flint},
  {Michigan}.
\newblock In \textit{Proceedings of the 24th ACM SIGKDD International
  Conference on Knowledge Discovery \& Data Mining}, Association for Computing
  Machinery, New York, NY, USA, p. 5–14.
%Type = Article
\bibitem[{{Advocate}(2021)}]{advocate_2021}
{Advocate}, 2021.
\newblock City announces effort to increase removal of lead services lines in
  2022.
\newblock \textit{{The Malden Advocate.}}
\newblock
  \eprint{{https://publizr.com/advocatenewsma/malden-advocate-10-slash-15-slash-21?html=true\#/3/}}.
%Type = Article
\bibitem[{Beher et~al.(2016)Beher, Possingham, Hoobin, Dougall and
  Klein}]{beher_prioritising_2016}
Beher, J., Possingham, H.P., Hoobin, S., Dougall, C., Klein, C., 2016.
\newblock Prioritising catchment management projects to improve marine water
  quality.
\newblock \textit{Environmental Science \& Policy} 59, 35--43.
%Type = Article
\bibitem[{Christenson(2022)}]{christenson_2022}
Christenson, G., 2022.
\newblock {Mayor Christenson's 2022 State of the City Address}.
\newblock \textit{{Malden Access Television,}}
\newblock {via YouTube (at timestamp 31:25)},
  \eprint{https://youtu.be/4hUeEGPewZ8?t=1885}.
%Type = Article
\bibitem[{{Clean Water Action}(2019)}]{clean_water_action_2019}
{Clean Water Action}, 2019.
\newblock {Chelsea} {Leaders} {Receive} {MWRA} {Grant} for {Removing} {Lead}
  {Service} {Lines}.
\newblock \textit{LSLR Collaborative.}
%Type = Article
\bibitem[{Costa et~al.(2014)Costa, Murta and Ribeiro}]{costa_applying_2014}
Costa, F., Murta, L., Ribeiro, C.C., 2014.
\newblock Applying software engineering techniques in the development and
  management of linear and integer programming applications.
\newblock \textit{International Transactions in Operational Research} 21, 6,
  1001--1030.
%Type = Article
\bibitem[{Dantzig(1957)}]{dantzig_discrete-variable_1957}
Dantzig, G.B., 1957.
\newblock Discrete-variable extremum problems.
\newblock \textit{Operations Research} 5, 2, 266--288.
%Type = Article
\bibitem[{Eiselt and Marianov(2014)}]{eiselt_multicriteria_2014}
Eiselt, H., Marianov, V., 2014.
\newblock Multicriteria decision making under uncertainty: a visual approach.
\newblock \textit{International Transactions in Operational Research} 21, 4,
  525--540.
%Type = Article
\bibitem[{Gabrilska(2021)}]{gabrilska}
Gabrilska, M., 2021.
\newblock Malden uses data to prioritize removal of lead-lined service pipes.
\newblock \textit{{Massachusetts Municipal Association (MMA).}}
\newblock
  \eprint{https://www.mma.org/malden-uses-data-to-prioritize-removal-of-lead-lined-service-pipes/}.
%Type = Incollection
\bibitem[{Goldberg(2017)}]{goldberg_2017}
Goldberg, D.W., 2017.
\newblock Geocoding.
\newblock In Richardson, D., Castree, N., Goodchild, M.F., Kobayashi, A., Liu,
  W. and Marston, R.A. (eds.), \textit{International Encyclopedia of Geography:
  People, the Earth, Environment and Technology}.
\newblock Wiley.
%Type = Article
\bibitem[{Goovaerts(2017)}]{goovaerts_2017}
Goovaerts, P., 2017.
\newblock How geostatistics can help you find lead and galvanized water service
  lines: {The} case of {Flint}, {MI}.
\newblock \textit{Science of The Total Environment} 599-600, 1552--1563.
%Type = Article
\bibitem[{Hu et~al.(2007)Hu, Shih, Rothenberg and Schwartz}]{hu_2007}
Hu, H., Shih, R., Rothenberg, S., Schwartz, B.S., 2007.
\newblock The epidemiology of lead toxicity in adults: Measuring dose and
  consideration of other methodologic issues.
\newblock \textit{Environmental Health Perspectives} 115, 3, 455–462.
%Type = Article
\bibitem[{Joseph et~al.(2009)Joseph, Maloney and
  Possingham}]{joseph_optimal_2009}
Joseph, L.N., Maloney, R.F., Possingham, H.P., 2009.
\newblock Optimal allocation of resources among threatened species: a project
  prioritization protocol.
\newblock \textit{Conservation Biology} 23, 2, 328--338.
%Type = Book
\bibitem[{Kellerer et~al.(2004)Kellerer, Pferschy and
  Pisinger}]{kellerer_knapsack_2004}
Kellerer, H., Pferschy, U., Pisinger, D., 2004.
\newblock \textit{Knapsack Problems}.
\newblock Springer.
%Type = Article
\bibitem[{Koç et~al.(2009)Koç, Morton, Popova, Hess, Kee and
  Richards}]{koc_prioritizing_2009}
Koç, A., Morton, D.P., Popova, E., Hess, S.M., Kee, E., Richards, D., 2009.
\newblock Prioritizing project selection.
\newblock \textit{The Engineering Economist} 54, 4, 267--297.
%Type = Article
\bibitem[{Lockitch(1993)}]{lockitch_1993}
Lockitch, G., 1993.
\newblock Perspectives on lead toxicity.
\newblock \textit{Clinical Biochemistry} 26, 5, 371–381.
%Type = Book
\bibitem[{Martello and Toth(1990)}]{martello_knapsack_1990}
Martello, S., Toth, P., 1990.
\newblock \textit{Knapsack problems: algorithms and computer implementations}.
\newblock John Wiley \& Sons.
%Type = Article
\bibitem[{Maturana and Eterovic(1995)}]{maturana_vehicle_1995}
Maturana, S., Eterovic, Y., 1995.
\newblock Vehicle routing and production planning decision support systems:
  Designing graphical user interfaces.
\newblock \textit{International Transactions in Operational Research} 2, 3,
  233.
%Type = Article
\bibitem[{O'Malley(2017)}]{omalley_2017}
O'Malley, R., 2017.
\newblock Lead pipes: The cost of kicking the can down the road.
\newblock \textit{{Public Health Post.}}
\newblock
  \eprint{https://www.publichealthpost.org/viewpoints/the-cost-of-lead-pipes/}.
%Type = Techreport
\bibitem[{Pannell(2015)}]{pannell_ranking_2015}
Pannell, D.J., 2015.
\newblock {Ranking Projects for Water-Sensitive Cities}.
\newblock Working Papers 204263, University of Western Australia, School of
  Agricultural and Resource Economics.
%Type = Article
\bibitem[{Pannell and Gibson(2016)}]{pannell_gibson_2016}
Pannell, D.J., Gibson, F.L., 2016.
\newblock Environmental cost of using poor decision metrics to prioritize
  environmental projects: Cost of poor decision metrics.
\newblock \textit{Conservation Biology} 30, 2, 382--391.
%Type = Article
\bibitem[{Potash et~al.(2020)Potash, Ghani, Walsh, Jorgensen, Lohff, Prachand
  and Mansour}]{potash_2020}
Potash, E., Ghani, R., Walsh, J., Jorgensen, E., Lohff, C., Prachand, N.,
  Mansour, R., 2020.
\newblock Validation of a machine learning model to predict childhood lead
  poisoning.
\newblock \textit{JAMA Network Open} 3, 9, e2012734.
%Type = Article
\bibitem[{Rocheleau(2016)}]{globe_2016}
Rocheleau, M., 2016.
\newblock Lead water pipes still a concern in boston area.
\newblock \textit{{The Boston Globe.}}
\newblock \eprint{https://archive.today/XlIxm}.
%Type = Article
\bibitem[{Sahni(1975)}]{sahni_1975}
Sahni, S., 1975.
\newblock Approximate algorithms for the 0/1 knapsack problem.
\newblock \textit{Journal of the ACM} 22, 1, 115–124.
%Type = Article
\bibitem[{Senju and Toyoda(1968)}]{senju_approach_1968}
Senju, S., Toyoda, Y., 1968.
\newblock An approach to linear programming with 0–1 variables.
\newblock \textit{Management Science} 15, 4, B--196.
%Type = Article
\bibitem[{Trueman et~al.(2016)Trueman, Camara and Gagnon}]{trueman_2016}
Trueman, B.F., Camara, E., Gagnon, G.A., 2016.
\newblock Evaluating the effects of full and partial lead service line
  replacement on lead levels in drinking water.
\newblock \textit{Environmental Science \& Technology} 50, 14, 7389--7396.
%Type = Techreport
\bibitem[{{US ATSDR}(2020)}]{hhs_2020}
{US ATSDR}, 2020.
\newblock {Toxicological Profile for Lead}.
\newblock Technical report CAS \#7439-92-1, {Agency for Toxic Substances and
  Disease Registry (ATSDR), US Department of Health and Human Services}.
%Type = Techreport
\bibitem[{{US CDC}(2012)}]{cdc_2012}
{US CDC}, 2012.
\newblock Low Level Lead Exposure Harms Children: A Renewed Call for Primary
  Prevention.
\newblock Committee report, {Advisory Committee on Childhood Lead Poisoning
  Prevention, US Centers for Disease Control and Prevention}.
%Type = Article
\bibitem[{{US Census}(2021a)}]{census_cps}
{US Census}, 2021a.
\newblock America’s families and living arrangements: 2021.
\newblock \textit{{U.S.~Census Bureau.}}
%Type = Article
\bibitem[{{US Census}(2021b)}]{census_tiger}
{US Census}, 2021b.
\newblock Tiger line shapefiles and tiger line files technical documentation.
\newblock \textit{{U.S.~Census Bureau.}}
%Type = Article
\bibitem[{{US EPA}(2021)}]{epa_lcrr}
{US EPA}, 2021.
\newblock Stronger protections from lead in drinking water: Next steps for the
  lead and copper rule.
\newblock \textit{{U.S.~Environmental Protection Agency.}}
%Type = Incollection
\bibitem[{Whitmyre and Driver(2005)}]{whitmyre}
Whitmyre, G., Driver, J., 2005.
\newblock Exposure assessment.
\newblock In Wexler, P. (ed.), \textit{Encyclopedia of Toxicology}.
\newblock Elsevier,  (2 edn.), p. 303–306.
%Type = Article
\bibitem[{Zartarian et~al.(2017)Zartarian, Xue, Tornero-Velez and
  Brown}]{zartarian_2017}
Zartarian, V., Xue, J., Tornero-Velez, R., Brown, J., 2017.
\newblock Children’s lead exposure: A multimedia modeling analysis to guide
  public health decision-making.
\newblock \textit{Environmental Health Perspectives} 125, 9.

\end{thebibliography}

\newpage

\appendix

\section*{Appendix A}
\label{appendix:payload}

\subsection*{Example response from the Google Maps Platform Geocoding web service}

The following output is the result of geocoding the text string \texttt{"1 mapel st malden ma"} (note the intentional misspelling of ``Maple'') using Google's geocoding \href{https://developers.google.com/maps/documentation/geocoding/overview}{web service} which is returned in JavaScript object notation (JSON) format:

\footnotesize
\begin{minted}{json}
    [
      {
        "address_components": [
          {
            "long_name": "1", "short_name": "1",
            "types": ["street_number"]
          },
          {
            "long_name": "Maple Street", "short_name": "Maple St",
            "types": ["route"]
          },
          {
            "long_name": "Malden", "short_name": "Malden",
            "types": ["locality", "political"]},
          {
            "long_name": "Middlesex County", "short_name": "Middlesex County",
            "types": ["administrative_area_level_2", "political"]
          },
          {
            "long_name": "Massachusetts", "short_name": "MA",
            "types": ["administrative_area_level_1", "political"]
          },
          {
            "long_name": "United States", "short_name": "US",
            "types": ["country", "political"]
          },
          {
            "long_name": "02148", "short_name": "02148",
            "types": ["postal_code"]
          }
        ],
        "formatted_address": "1 Maple St, Malden, MA 02148, USA",
        "geometry": {
          "location": {"lat": 42.4293921, "lng": -71.07431389999999},
          "location_type": "ROOFTOP",
          "viewport": {
            "northeast": {"lat": 42.4307410802915, "lng": -71.07296491970848},
            "southwest": {"lat": 42.4280431197085, "lng": -71.0756628802915}
          }
        },
        "place_id": "ChIJX7ItxV9x44kREQUjrcSGqcI",
        "plus_code": {
          "compound_code": "CWHG+Q7 Malden, MA, USA",
          "global_code": "87JCCWHG+Q7"
        },
        "types": ["street_address"]
      }
    ]
\end{minted}

\end{document}